\documentclass{PoS}
\usepackage{amsmath}
\usepackage{wasysym}
\usepackage{multirow}
\title{Bottom Quark Mass from $\Upsilon$ Sum Rules to  ${\cal O}(\alpha_s^3)$}

\ShortTitle{Bottom Quark Mass from $\Upsilon$ Sum Rules to  ${\cal O}(\alpha_s^3)$}

\author{\speaker{Nikolai Zerf}\\
        University of Alberta\\
        E-mail: \email{zerf@ualberta.ca}}

\abstract{We present the deterimination of the bottom quark mass using non-relativistic $\Upsilon$ Sum Rules at $\text{N}^3\text{LO}^*$~\cite{Penin:2014zaa}.
 The explicit dependence of $\overline{m}_b(\overline{m}_b)$ on the input value $\alpha_s(M_Z)$ is given for the first time.
}

\FullConference{ Loops and Legs in Quantum Field Theory - LL 2014,\\
		 27 April - 2 May 2014 \\
		 Weimar, Germany }

\begin{document}

\section{Introduction}
The  bottom quark mass $m_b$ is one of the fundamental parameters of the Standard Model (SM). 
It cannot be predicted from first principles with the current understanding of fundamental physics.
In order to determine its value we are currently only able to match calculable observables to their experimental measured counterparts.
Here the mass is taken as free parameter and chosen in such a way, that the theory prediction agrees with the experimental result.
This means we are extracting the bottom quark mass from experimental data.

It is important to have highest possible precision of the extracted mass value,
because it enters any prediction where the bottom quark mass appears.
Thus gaining a more accurate result for $m_b$ results in an improvement of all theory predictions incorporating the bottom quark mass at once.
The motivation to improve the accuracy of $m_b$ is originating from many different fields of today's particle physics like flavor physics, top physics, collider physics,
theories beyond SM like Grand Unified Theories with or without Super Symmetry.

\section{Overview}
In this section we give a very short overview of extraction methods used to obtain the bottom quark mass from experimental $e^+e^-\rightarrow b \overline{b}$ data.
For all the methods displayed below one requires the measured values of the $R$-Ratio in dependence of the center of mass energy $\sqrt{s}$ of the incomming $e^+e^-$-pair, 
which is defined as ratio of two production cross sections:
\begin{equation}
R_{exp}(s) = {1\over Q_b^2}{\sigma(e^+e^-\rightarrow {b\bar b})\over
\sigma(e^+e^-\rightarrow \mu^+\mu^-)} \,,
\label{eq::spectralexp}
\end{equation}
The extraction methods are sensitive to different $\sqrt{s}$-regions:\\
$\bullet$ \emph{spectroscopy based on perturbation theory}: Requires the peak position of the narrow bound state resonances $\Upsilon_{1 S}$ below threshold ($\sqrt{s}< 2 m_b$).
 With the measured value of the resonance mass $M_{\Upsilon(1 S)}$ one can extract $m_b$ once the binding energy $E_{1}$ has been calculated in dependence of $m_b$ on theory side:
 \begin{equation}
  M_{\Upsilon(1 S)}=2 m_b+E_{1}
 \end{equation}
 Apart from the simple extraction equation, 
 one misses sizable non perturbative (n.p.) contributions to $E_1$,
 which grow rapidly with increasing $n$ value.
 This restricts the application of this method to the first resonance.
 Otherwise the unknown systematic shift made up by n.p. contributions becomes too large.
 However, on perturbation theory (p.t.) side one has to face a minimal amount of calculations compared to the other p.t. methods.\\
$\bullet$ \emph{spectroscopy based on lattice simulations}:
To be able to account for the n.p. contributions to the binding energies, 
one can use lattice simulations to determine the binding energy.
For example one can use Non-Relativistic QCD (NRQCD) as effective field theory in order
to encode high energetic degrees of freedom ($\sim m_b$) into Wilson (or matching) coefficients, which appear in the NRQCD Lagrangian.
Once all required coefficients have been calculated within perturbative lattice QCD (pLQCD) up to sufficient high order,
a full lattice simulation is needed to determine the binding energy of the $b\overline{b}$-system.
For more details concerning this method, the reader may be referred to Ref.~\cite{Lee:2013mla}.\\
$\bullet$ \emph{Relativistic $\Upsilon$ Sum Rules}:
In order to reduce the impact of n.p. effects in the extraction of $m_b$ one may construct a more ``inclusive'' observable from $R(s)$.
in case of Relativistic $\Upsilon$ Sum Rules the observable is given by the $n$-th moment:
\begin{equation}
{\cal M}_n=
(4m_b^2)^n\int_0^\infty{R(s)ds\over s^{n+1}}=\left.{12\pi^2\over n!}(4m_b^2)^n{d^n\over ds^n}\Pi(s)\right|_{s=0}\,.
\label{eq::momentdef}
\end{equation}
Here the weighted integral over $R(s)$ can be rewritten using the optical theorem $R(s)=12\pi {\rm Im}\Pi(s+i\epsilon)$ and the analytic properties of the vacuum polarization function $\Pi(s)$
\begin{equation}
\left(q_\mu q_\nu-g_{\mu \nu}q^2\right)\Pi(q^2)=
i\int d^dx\,e^{iqx}\,\langle 0|Tj_{\mu}(x)j_{\nu}(0)|0\rangle \,.
\label{eq::vacuumpol}
\end{equation}
defined through the two-point vacuum correlator of the  heavy-quark electromagnetic current $j_\mu=\bar b\gamma_\mu b$.
So the $n$-th moment is defined (up to some prefactor) by the $n$-th derivative of vacuum polarization function at $s=0$,
which is far away from any physical state affected by n.p. contribution and thus calculable reliably within p.t..
That means we can expand the vacuum polarization function in powers of the strong coupling constant $\alpha_s$
when we are interested in the region where $z=\frac{s}{4m_b^2}$ is close to zero:
\begin{equation}
 \Pi(s)=\sum_{i}\sum\limits_{n\geq0}\left(\frac{\alpha_s}{4\pi}\right)^i C_{i,n}z^n\,.  
\end{equation}
For the $n$-th moment one needs the coefficients $C_{i,n}$ with all $i$ less or equal to the loop order we are interested in.
To extract the bottom quark mass, one now requires that $n$-th experimental moment is in agreement with the $n$-th theory moment:
\begin{equation}
{\cal M}^{exp}_n = {\cal M}^{th}_n\,,
\label{eq::sumrules}
\end{equation}
The knowledge of the $C_{4,n}$\footnote{is known up to including $n=4$ analytically and up to including $n=10$ numerically} enabled the authors of Ref.~\cite{Chetyrkin:2010ic} to extract the bottom quark mass with -- to that date -- highest precision,
although there is one minor drawback when using this extraction method:
Because one can only use low moment numbers $n=1,\dots, 4$ (like explained below), 
the integral contribution in Eq.~(\ref{eq::momentdef}) around the threshold region is not sufficient suppressed so it cannot be dropped when calculating the moment on the experimental side.
That means one has to deal with the far less accurate data (compared to the precisely measured resonances) for $R_{exp}(s\approx 4m_b^2)$ above threshold.\\
$\bullet$ \emph{Non-Relativistic $\Upsilon$ Sum Rules}:
In order to avoid the experimental data above threshold one suppresses the corresponding $\sqrt{s}$ region.
Therefore one just has to choose higher moments $n=10,\dots,20$ for the extraction.
However, in this setup one cannot rely on the fixed order p.t. prediction given by $C_{i,n}$~\cite{Voloshin:1987rp},
because the moment integral receives dominant contributions from the region where the $b\overline{b}$ pair has a non-relativistic relative velocity $v$
and the characteristic scaling $\alpha_s \sim v \sim \frac{1}{\sqrt{n}}$ makes a resummation of all $k$-Coulomb-gluon exchanges $\sim (\frac{\alpha_s}{v})^k\sim1$ between the $b\overline{b}$ pair mandatory.
Radiative corrections on top of this Coulomb-resummation can be implemented within the framework of the effective field theory Potential Non-Relativistic QCD (pNRQCD).
For a recent comprehensive review of this theory and the required matching coefficients the reader may be referred to Ref.~\cite{Beneke:2013jia} and references therein.
In order to calculate the $n$-moment for higher values of $n$, one has to calculate the vacuum polarization function $\Pi(s)$ within pNRQCD up to the required order ($\text{N}^3\text{LO}$).
$R(s)$ can then be obtained via the optical theorem.

Once $R(s)$ has be calculated within pNRQCD one can choose higher moment values $n$ to extract the bottom quark mass and
one can just use the precisely measured leptonic widths $\Gamma_{\Upsilon(mS)\to l^+l^-}$ and resonance masses $M_{\Upsilon(mS)}$ of the bounds states only.
The experimental moment is then given by:
\begin{equation}
{\cal M}^{exp}_n={(4m_b^2)^n}{9\pi\over  Q_b^2\alpha^2(2m_b)}\left(
\sum_{m}{\Gamma_{\Upsilon(mS)\to l^+l^-}\over M_{\Upsilon(mS)}^{2n+1}}
\right)\,,
\label{eq::momentexp}
\end{equation}
Here $Q_b^2$ is the charge of the bottom quark and  $\alpha(2m_b)$ is the running fine structure constant at the scale $\mu=2m_b$.


\section{Bottom quark mass extraction using Non-Relativistic $\Upsilon$ Sum Rules}
The vacuum polarization $\Pi(s)$ within pNRQCD can be cast in the following form ($E=\sqrt{s}-2m_b$):
\begin{equation}
\Pi(s)={N_c\over 2 m_b^2}\left(c_v-{E\over m_b}{d_v\over 6}+\ldots\right)^2
\left(1+{E\over 2m_b}\right)^{-2} G^{s}(0,0;E)\,,
\label{eq::vacuumpolnr}
\end{equation}
Here $c_{v}$ and $d_{v}$ are the NRQCD matching coefficient of the heavy quark vector current and 
$G^{s}(0,0;E)$ is the Green's function governing the time evolution of the non-relativistic $b\overline{b}$-pair including radiative corrections.
$N_c=3$ is the number of colors.
Because the spectral property of the Green's function allows to decomposes the latter
into bound state contributions (build up by the wave functions $|\psi_n(0)|^2$ taken at the origin and the corresponding energy eigenvalue $E_n$) below threshold 
\begin{equation}
G^s(0,0;E)=\sum_{n=1}^\infty{|\psi_n(0)|^2\over
  E_n-E-i\epsilon} +G^s_{\text{cont}}(0,0;E)\,,
\label{eq::spectralrep}
\end{equation}
and continuum contributions above threshold,
the $n-$th moment splits into two separate pieces:
\begin{align}
{\cal M}_n&=(4m_b^2)^n\left(
{48\pi^2N_c}\sum_{m}{ C_{v,m}^2|\psi_m(0)|^2\over (2m_b+E_m)^{2n+3}}
+\int_{4m_b^2}^\infty{R(s)ds\over s^{n+1}}\right)\,, &  C_{v,m}&=c_v-{E_m\over m_b}{d_v\over 6}\,.
\label{eq::momentnr}
\end{align}
A discrete sum of resonance contributions below threshold and an integral forming the continuum contribution above threshold.
In order to calculate ${\cal M}_n$ at  $\text{N}^3\text{LO}$ one needs the following building blocks up to this order:
 {(i)}  \emph{$E_n$}  \cite{Pineda:1997hz,Penin:2002zv,Penin:2005eu,Beneke:2005hg},
 {(ii)}  \emph{$|\psi_n(0)|^2$}  \cite{Penin:2005eu,Beneke:2005hg,Beneke:2007gj,Beneke:2007pj,Beneke:2014qea},
 {(iii)} \emph{$d_v$}  \cite{Luke:1997ys},
 {(iv)} \emph{$c_v$} available up to small fermionic singlet contribution  \cite{Marquard:2006qi,Marquard:2009bj,Marquard:2014pea},
 {(v)} \emph{$G^s_{\text{cont}}(0,0;E)$} not fully available.\\
Because the continuum contribution of the Green's function is not fully available at $\text{N}^3\text{LO}$ and its contribution is suppresed for larger moments $n$, 
we use the following approximation above threshold:
\begin{equation}
 R^{\text{N}^3\text{LO}}(s)\approx \rho  \frac{R^{\text{N}^3\text{LO}}(s)}{R^{\text{NNLO}}(s)}\Big|_{\sqrt{s}=2m_b+E_1}R^{\text{NNLO}}(s)\,.
\end{equation}
Where the induced variation by $\frac{1}{2}\leq\rho\leq2$ in the mass is covered by the error estimate $\Delta_\rho$.

\section{${\rm OS}\rightarrow\overline{\rm MS}$ transition and error estimate}
After the extraction of the on-shell mass $m_b$ in dependence of the renormalization scale $\mu$ and moment number $n$, 
we cancel the renormalons by switching to the short distance $\overline{\rm MS}$-mass $\overline{m}_b(\overline{m}_b)$:
\begin{equation}
  m^{\text{N}^m\text{LO}}_b(\mu)= \overline{m}_b(\overline{m}_b) r^{\text{N}^{m+1}\text{LO}}(\mu,\,\overline{m}_b(\overline{m}_b))\,.
\end{equation}
This requires the four loop ${\rm OS}\leftrightarrow\overline{\rm MS}$ conversion formula for the mass,
where currently the constant term is unknown.
By default we use for the latter the renormalon improved approximation given in Ref.~\cite{Pineda:2001zq} or the large-$\beta_0$ approximation.\\
$\bullet$ The resulting difference in the bottom quark mass between both approximations is used as an error estimate for the unknown constant term $\Delta_{r^{(4)}} $.\\
$\bullet$ In order to estimate the error $\Delta_{n.p.}$ arising from n.p. contribution we use the gluon condensate approximation~\cite{Voloshin:1995sf}.\\
$\bullet$ The error estimate of neglected higher order terms in the p.t. $\Delta_{p.t.}$ is defined by half of the observed shift in the mass when going from $\text{NNLO}$ to $\text{N}^3\text{LO}$.\\
$\bullet$ We also assign a $\Delta_n$ error which is given by $\tfrac{1}{2}|\overline{m}_b(\overline{m}_b,n=20)-\overline{m}_b(\overline{m}_b,n=10)|$,
where we extract the central value at $n=15$ and $\mu=\overline{m}_b(\overline{m}_b)$.\\
$\bullet$ The experimental error $\Delta_{exp}$ is estimated by the coherent variation of the experimental resonance data within the given uncertainties.\\
$\bullet$ The absolute uncertainty $\Delta \alpha_s(M_Z)$ in the value of $\alpha_s(M_Z)=\overline{\alpha_s}(M_Z)\pm\Delta \alpha_s(M_Z)=0.1184\pm 0.0007$ is kept fixed and induces
a variation in $\overline{m}_b(\overline{m}_b)$, which is covered by the error $\Delta_{\alpha_s}$.

\section{Charm mass effects}
Up to this point the charm quark has been treated as massless in all formulae and building blocks above.
However, due to the fact that it has a non-vanishing mass (we use $\overline{m}_c(\overline{m}_c)\approx 1.3\text{GeV}$) 
a shift of about $-25\pm 5 \text{MeV}$ in $\overline{m}_b(\overline{m}_b)$ is induced when applying the result of Ref.~\cite{Hoang:2000fm} to our values.
The additional induced uncertainty in $\overline{m}_b(\overline{m}_b)$ is called $\Delta_{m_c}$.

\section{Results}

\begin{table}
 \begin{center}
    \begin{tabular}{|c||c|c||c||c|c|c|c|c|c||c|}
     \hline
     $\overline{\alpha}_s(M_Z)$  & $\overline{m}_b(\overline{m}_b)|_{m_c=0}$ & $\overline{m}_b(\overline{m}_b)$ & $\Delta_{\alpha_s}$ & $\Delta_\rho $ &$\Delta_{r^{(4)}} $& $\Delta_{n}$ & $\Delta_{p.t.}$& $\Delta_{n.p.}$                 & $\Delta_{m_c} $& $\Delta_{exp}$ \\
     \hline
     $0.1200$                    & $4,190$                                   & $4,165$                          & $1.8$               & $3.2$          & $2.4$             & $2.1$        & $2.3$          & \multirow{3}{*}{$0.8$}          & \multirow{3}{*}{$5.0$}& \multirow{3}{*}{$2.3$}\\
     \cline{1-8}
     $0.1184$                    & $4,194$                                   & $4,169$                          & $1.9$               & $4.2$          & $2.2$             & $3.4$        & $2.1$          &                                 &                       &\\
     \cline{1-8}
     $0.1160$                    & $4,200$                                   & $4,175$                          & $1.9$               & $6.1$          & $1.8$             & $5.5$        & $1.7$          &                                 &                       &\\
     \cline{1-8}
     $0.1130$                    & $4,208$                                   & $4,183$                          & $1.9$               & $9.1$          & $1.5$             & $8.1$        & $1.3$          &                                 &                       &\\
     \hline
    \end{tabular}

     \caption{$\overline{m}_b(\overline{m}_b)$ at $\text{N}^3\text{LO}^*$ in dependence of central value $\overline{\alpha}_s(M_Z)$ and different contributions to the uncertainty of it.
     All units are given by MeV.\label{tab1}}
  \end{center}
\end{table}
In Table~\ref{tab1} we list the extracted values for $\overline{m}_b(\overline{m}_b)$ at $\text{N}^3\text{LO}^*$ for vanishing and non-vanishing charm quark mass and the associated uncertainties in dependence of the chosen central value $\overline{\alpha}_s(M_Z)$.
Choosing the results of the second row, leads to the published result of Ref.~\cite{Penin:2014zaa} after adding up the theory errors stated in column five to ten in quadrature:
\begin{equation}
\overline{m}_b(\overline{m}_b)=4.169\pm 0.008_{th}\pm 0.002_{\alpha_s}\pm 0.002_{exp} \,.
\label{eq::final}
\end{equation}
This value is in good agreement with the result obtained using relativistic sum rules~\cite{Chetyrkin:2010ic}.
For lower central values of $\alpha_s(M_Z)$ we observe a stronger dependence of the result on the moment number $n$.
But because our continuum approximation becomes less accurate there too, 
we are not in the position to conclude that the data disfavors lower values of $\alpha_s$ at $\text{N}^3\text{LO}$.
This, however, is true at $\text{NNLO}$
.

Further we observe that the $\text{N}^3\text{LO}^*$ result is far more stable
than the $\text{NNLO}$ result in the given parameter space
which allows for a significant reduction of the theoretical uncertainty.


\begin{thebibliography}{99}

\bibitem{Penin:2014zaa}
  A.~A.~Penin and N.~Zerf,
  JHEP {\bf 1404} (2014) 120
  [arXiv:1401.7035 [hep-ph]].


\bibitem{Lee:2013mla}
  A.~J.~Lee {\it et al.}  [HPQCD Collaboration],
  Phys.\ Rev.\ D {\bf 87} (2013) 7,  074018
  [arXiv:1302.3739 [hep-lat]].
\bibitem{Chetyrkin:2010ic}
  K.~Chetyrkin, J.~H.~Kuhn, A.~Maier, P.~Maierhofer, P.~Marquard, M.~Steinhauser and C.~Sturm,
  Theor.\ Math.\ Phys.\  {\bf 170} (2012) 217
  [arXiv:1010.6157 [hep-ph]].
 
\bibitem{Voloshin:1987rp}
  M.~B.~Voloshin and Y.~.M.~Zaitsev,
  Sov.\ Phys.\ Usp.\  {\bf 30} (1987) 553
   [Usp.\ Fiz.\ Nauk {\bf 152} (1987) 361].
 
\bibitem{Beneke:2013jia}
  M.~Beneke, Y.~Kiyo and K.~Schuller,
  arXiv:1312.4791 [hep-ph].

 
\bibitem{Pineda:1997hz}
  A.~Pineda and F.~J.~Yndurain,
  Phys.\ Rev.\ D {\bf 58} (1998) 094022
  [hep-ph/9711287].

\bibitem{Penin:2002zv}
  A.~A.~Penin and M.~Steinhauser,
  Phys.\ Lett.\ B {\bf 538}, 335 (2002)
  [hep-ph/0204290].

\bibitem{Penin:2005eu}
  A.~A.~Penin, V.~A.~Smirnov, and M.~Steinhauser,
  Nucl.\ Phys.\  B {\bf 716 } (2005)  303
  [hep-ph/0501042].

\bibitem{Beneke:2005hg}
  M.~Beneke, Y.~Kiyo, and K.~Schuller,
  Nucl.\ Phys.\  B {\bf 714 } (2005)  67
  [hep-ph/0501289].

\bibitem{Beneke:2007gj}
  M.~Beneke, Y.~Kiyo, and K.~Schuller,
  Phys.\ Lett.\  B {\bf 658 } (2008)  222
  [arXiv:0705.4518 [hep-ph]].

\bibitem{Beneke:2007pj}
  M.~Beneke, Y.~Kiyo, and A.~A.~Penin,
  Phys.\ Lett.\  B {\bf 653 } (2007)  53
  [arXiv:0706.2733 [hep-ph]].

\bibitem{Beneke:2014qea}
  M.~Beneke, Y.~Kiyo, P.~Marquard, A.~Penin, J.~Piclum, D.~Seidel and M.~Steinhauser,
  Phys.\ Rev.\ Lett.\  {\bf 112}, 151801 (2014)
  [arXiv:1401.3005 [hep-ph]].
  
\bibitem{Luke:1997ys}
  M.~E.~Luke and M.~J.~Savage,
  Phys.\ Rev.\ D {\bf 57} (1998) 413
  [hep-ph/9707313].

\bibitem{Marquard:2006qi}
  P.~Marquard, J.~H.~Piclum, D.~Seidel, and M.~Steinhauser,
  Nucl.\ Phys.\  B {\bf 758 } (2006)  144
  [hep-ph/0607168].

\bibitem{Marquard:2009bj}
  P.~Marquard, J.~H.~Piclum, D.~Seidel, and M.~Steinhauser,
  Phys.\ Lett.\  B {\bf 678 } (2009)  269
  [arXiv:0904.0920 [hep-ph]].

\bibitem{Marquard:2014pea}
  P.~Marquard, J.~H.~Piclum, D.~Seidel and M.~Steinhauser,
  arXiv:1401.3004 [hep-ph].


\bibitem{Pineda:2001zq}
  A.~Pineda,
  JHEP {\bf 0106} (2001) 022
  [hep-ph/0105008].

\bibitem{Voloshin:1995sf}
  M.~B.~Voloshin,
  Int.\ J.\ Mod.\ Phys.\ A {\bf 10} (1995) 2865
  [hep-ph/9502224].

\bibitem{Hoang:2000fm}
  A.~H.~Hoang,
  hep-ph/0008102.

  
  
\end{thebibliography}
\end{document}